\documentclass[aps,prl,twocolumn,showpacs,preprintnumbers]{revtex4}
\usepackage{amsmath,amssymb}
\usepackage{graphicx}

\newcommand\weber{\mathrm{weber}}

\newcommand\sheet{\mathrm{sheet}}
\newcommand\ext{\mathrm{ext}}
\newcommand\ind{\mathrm{ind}}

\newcommand\bfA{\mathbf{A}}
\newcommand\bfB{\mathbf{B}}
\newcommand\bfJ{\mathbf{J}}
\newcommand\bfr{\mathbf{r}}
\newcommand\bfz{\mathbf{z}}
\newcommand\bfnabla{\boldsymbol{\nabla}}

\begin{document}

\title{Quantum-based Mechanical Force Realization in Pico-Newton  Range}

\author{Jae-Hyuk Choi}
\email{jhchoi@kriss.re.kr}
\affiliation{Mechanical Metrology Group, Division of Physical Metrology, KRISS, Korea}
\author{Min-Seok Kim}
\affiliation{Mechanical Metrology Group, Division of Physical Metrology, KRISS, Korea}
\author{Yon-Kyu Park}
\affiliation{Mechanical Metrology Group, Division of Physical Metrology, KRISS, Korea}

\author{Mahn-Soo Choi}
\email{choims@korea.ac.kr}
\affiliation{Department of Physics, Korea University, Seoul 136-701, Korea}

\date{\today}

\begin{abstract}
We propose mechanical force realization based on flux quantization in
the pico-Newton range.  By controlling the number of flux quantum in a
superconducting annulus, a force can be created as integer multiples of
a constant step.  For a 50 nm-thick Nb annulus with the inner and outer
radii of 5 $\mu$m and 10 $\mu$m, respectively, and the field gradient of
10 T/m the force step is estimated to be 184 fN.
The stability against thermal fluctuations is also addressed.
\end{abstract}

\pacs{06.20.fb, 85.25.-j, 84.71.Ba}

\maketitle

\paragraph{Introduction}
Due to remarkable improvement in its sensitivity, force measurement has
become a useful and essential probe for leading-edge
nano/bio-researches\cite{Pratt04}, which cover nanoscale imaging,
protein folding studies, nanoindentation, and many others. The force
detection limit keeps getting lowered, for example, to an atto-Newton
(10$^{-18}$ N) level in magnetic resonance force microscopy.  Such an
ultimate resolution is enough to read a single electron
spin.\cite{Rugar04,Knobel03a,Huang03a}

Unfortunately, however, no direct SI-traceable force realization has
been established even at sub-Newton level, 
which means no standard to compare with measured force.
Prevailing dead-weight method, which creates gravitational force
using standard weights, obviously becomes no
longer valid below micro-Newton level.\cite{Pratt04}
At the micro-Newton level, an electrostatic force realization has been
proposed recently by Pratt \textit{et al}.  Therein, electrostatic
standard force is created between two coaxial electrodes in maintaining
a constant voltage and expected to allow a relative uncertainty of
parts in $10^4$.  Related electrical units are traced to their standards
based on Josephson and quantized Hall effects.
At the nano-Newton or pico-Newton level, no force realization for
standard has been suggested despite needs for precision measurements of
force.  It is essential, for instance, to testify fundamental forces
such as Casimir force, non-Newtonian gravitation,\cite{Chiave03} etc.

Also, it casts a striking contrast to the case of electrical units such
as voltage, that (to our best knowledge) no attempt has been tried to
directly use quantum phenomena in realizing a mechanical force.
Here, we present a concept of quantum-based force realization utilizing
a macroscopic quantum phenomenon known as magnetic flux quantization in
a superconducting annulus.  Magnetic force exerted on flux quanta can be
increased or decreased by a constant step, which is estimated to be
sub-pico-Newton level.  Moreover,
those force-generating states are stable and
robust against thermal fluctuation at liquid helium temperature.

\begin{figure}
\includegraphics*[width=8cm]{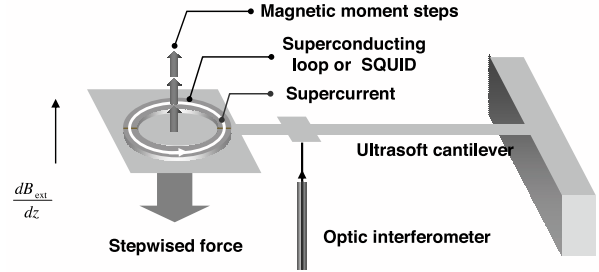}
\caption{A schematic of force realization based on flux quantization in
  superconducting annulus.}
\label{fig:schem}
\end{figure}

\paragraph{Basic Principles}
Figure~\ref{fig:schem} shows a schematic of quantum-based force realization. A
tens-micron-sized superconducting annulus is mounted on an ultrasoft
micro-cantilever. Below a superconducting transition temperature of the
superconducting material, magnetic fluxoid through the annulus is
quantized, and the resultant magnetic moment has a component with
constant steps, the number of which depends on the quantum number.
The step size  is determined by fundamental constants such as electron charge
and the length quantities of the annulus.

In a calibrated magnetic field gradient, dB/dz, a force is created on
the superconducting annulus, and therefore on the micro-cantilever. Through
a designed procedure of magnetic field and temperature,
we can leave trapped flux quanta as many as we want in zero external field,
while the field gradient is not zero.
The displacement of the cantilever is monitored by optic interferometer or
lever and can be used in calibrating its spring constant.

\paragraph{Magnetic Moment of a Superconducting Annulus}
For the accurate force realization, it is essential first to
evaluate the magnetic moment of the superconducting annulus.

We consider a superconducting annulus of inner and outer radii, $a$
and $b$, respectively, and thickness $d$ in a uniform external magnetic
field $\bfB_\ext=B_\ext\hat\bfz$.  In response to the external magnetic
field or the fluxoid trapped in the hole, currents $\bfJ$ are induced in
the annulus and the field is modified. The net magnetic vector
potential
\begin{math}
\bfA = \bfA_\ext + \bfA_\ind
\end{math}
consists of the external part $\bfA_\ext$ and the induced part $\bfA_\ind$.
$\bfJ$ and $\bfA_\ind$ are related the Biot-Savart law
\begin{equation}
\label{Brandt::eq:0c}
\bfA_\ind(\bfr) =
\int{d^3\bfr'}\frac{\mu_0\bfJ(r')}{4\pi|\bfr-\bfr'|} \,,
\end{equation}
where $\mu_0$ is the magnetic permeability.  On the other hand, within a
superconductor, the current is governed by the London equation
\begin{equation}
\label{Brandt::eq:1}
\bfJ = -\frac{1}{\mu_0\lambda_L^2}\left(\bfA +
  \frac{\Phi_0}{2\pi}\bfnabla\theta\right) \,,
\end{equation}
where $\theta$ is the phase of the superconducting order parameter,
$\lambda_L$ is the London penetration depth, and
$\Phi_0\equiv{h}/2|e|\approx 2.07\times 10^{-15}\weber$ is the flux
quantum.  Equations~(\ref{Brandt::eq:0c}) and (\ref{Brandt::eq:1}) give
an integral equation for $\bfJ$, from which one can get the magnetic
moment.  Given the geometry of the superconductor, one can greatly
simplify the integral equation by following \cite{Brandt04a}.

Since the current flows only in the thin film ($-d/2\leq z\leq d/2$,
$d\ll\lambda_L$), it is customary to define a sheet current such that
\begin{math}
\bfJ(x,y,z) \approx \bfJ_\sheet(x,y)\delta(z)
\end{math}
and to represent the physical quantities in the polar coordinates
$(r,\varphi,z)$ so that
\begin{math}
\bfJ_\sheet(x,y) = \bfJ_\sheet(r)\hat\varphi
\end{math},
\begin{math}
\bfA(x,y,0) = [A_\ext(r)+A_\ind(r)]\hat\varphi
\end{math}, and
\begin{math}
\bfnabla\theta(x,y,0) = (n/r)\hat\varphi
\end{math}, where the integer $n$ is the number of fluxes trapped in the
hole.  Then the London equation~(\ref{Brandt::eq:1}) reads as
\begin{equation}
\label{Brandt::eq:8}
\mu_0J_\sheet(r) = -\frac{1}{\Lambda}
\left[\frac{n\Phi_0}{2\pi r} + A_\ext(r) + A_\ind(r)\right]
\end{equation}
where $\Lambda\equiv\lambda_L^2/d$ is the effective penetration depth and
the Biot-Savart law~(\ref{Brandt::eq:0c}) is rewritten as
\begin{equation}
\label{Brandt::eq:13}
A_\ind(r) = \int_a^b{dr'}Q(r,r')\, \mu_0J_\sheet(r') \,,
\end{equation}
where
\begin{equation}
\label{Brandt::eq:14}
Q(r,r') = \int_0^\pi\frac{d\varphi}{2\pi}\;
\frac{\cos\varphi}{\sqrt{1-2(r/r')\cos\varphi + (r/r')^2}} \,.
\end{equation}
From the London equation~(\ref{Brandt::eq:8}), one identifies two
sources: one from the \emph{London fluxoid} \cite{London54a}
\begin{math}
D_1(r) \equiv {n\Phi_0}/{2\pi r}
\end{math}
and the other from the external field
\begin{math}
D_2(r) \equiv A_\ext(r)
\end{math}. %
Accordingly, we decompose the induced quantities, $J_\sheet$ and $A_\ind$,
into contributions from $D_1$ and $D_2$, respectively:
\begin{math}
J_\sheet = J_1 + J_2
\end{math} and
\begin{math}
A_\ind = A_1 + A_2
\end{math}.
Combining the two equations~(\ref{Brandt::eq:8}) and
(\ref{Brandt::eq:13}), one arrives at the integral equations
for $J_1$ and $J_2$
\begin{equation}
\label{Brandt::eq:17}
\int_a^b{dr'}\;Q(r,r')\,\mu_0J_\nu(r') + \Lambda\,\mu_0J_\nu(r)
= -D_\nu(r),
\end{equation}
respectively ($\nu=1,2$)\cite{Brandt04a,Fetter80b}.  The nice feature of
this form is that $a\leq r\leq b$, which facilitates pretty much the
numerical solution of the equation.
Once $J_\nu$ ($\nu=1,2$) are at hand,
the magnetic moment $m=m_1+m_2$ is given by
\begin{equation}
\label{eq:3}
m_\nu = \pi\int_a^b{dr}\;r^2J_\nu(r) \,.
\end{equation}
We have followed \cite{Brandt04a} to discretize the integral
equation~(\ref{Brandt::eq:17}) and get $J_\nu$ and $m_\nu$ numerically.

\begin{figure}
\centering
\includegraphics*[width=8cm]{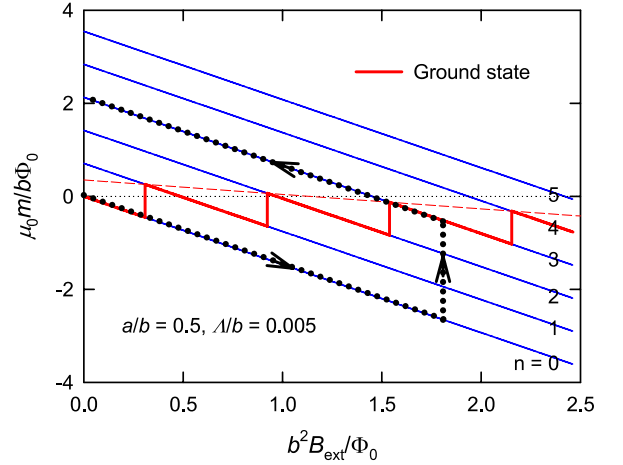}
\caption{(color online) Total magnetic moment as a function of $B_\ext$
  for $n$ trapped fluxoids (solid blue lines), $a/b=0.5$, and
  $\Lambda/b=0.005$.  The thick
  (red) line is for the ground state configuration.  The dashed (red)
  line is the guide for eyes connecting the peaks of ground-state
  magnetic moment. The thick dotted (black) line indicates the
  experimental procedure (see the text).}
\label{fig:moment}
\end{figure}

Figure~\ref{fig:moment} shows typical behavior of the total magnetic
moment $m$ of the superconducting annulus as a function of $B_\ext$.  It
is clearly seen that at a given $B_\ext$, the magnetic moment is
quantized depending on the number of trapped fluxes.

\begin{figure}
\centering
\includegraphics*[width=7cm]{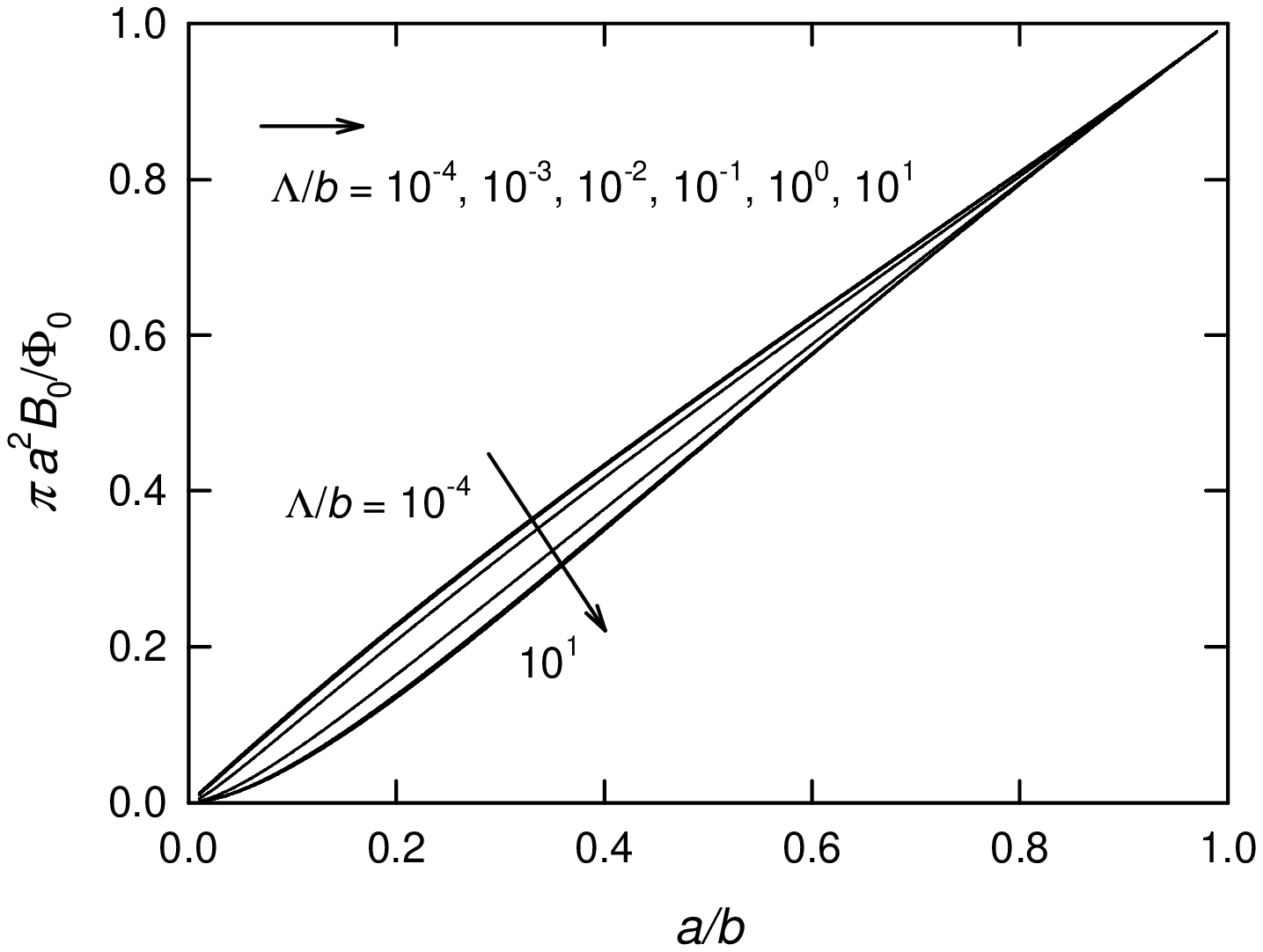}
\caption{$B_0$ as a function of $a/b$ for different values of $\Lambda/b$.}
\label{fig:period}
\end{figure}

At a given value of $B_\ext$, the allowed quantized value of the
magnetic moment corresponds to a metastable state of the system (the
thin blue lines in Fig.~\ref{fig:moment}).  The magnetic moment
corresponding to the most stable state (the thick red line in
Fig.~\ref{fig:moment}) is determined by the Gibbs free
energy \cite[]{Brandt04a}
\begin{equation}
\label{eq:6}
G_n(B_\ext) = -\frac{1}{2}m B_\ext + \frac{1}{2}n\Phi_0I \,,
\end{equation}
where $I=I_1+I_2$ with
\begin{math}
I_\nu = \int_a^b{dr}\; J_\nu(r)
\end{math}
is the total current along the annulus.
The ground-state magnetic moment is a periodic function of $B_\ext$ with
period $B_0$.  Figure~\ref{fig:period} plots $B_0$ versus $a/b$.

\paragraph{Estimation of Magnetic Moment, Force Step, Force Limit}
To estimate the magnitude of the magnetic moment, let us consider an Nb
annulus of inner and outer radii, $a = 5$ $\mu$m and $b = 10$ $\mu$m,
respectively, with thickness of 50 nm.  The Nb material was chosen
because its superconducting transition temperature, 9.25 K,\cite{Pool00}
is readily accessible in cryogenics and its thin film fabrication is
well established to have provided good quality of thin-film circuits
such as SQUID.  Its London penetration depth $\lambda_L$ and coherence
length $\xi$ are 50 nm and 39 nm, respectively.\cite{Pool00} The choice
of radii is reasonable considering the typical dimension of
microfabricated cantilevers with a low spring constant ($k = 10^{-4}$ to
$10^{-5}$ N/m).\cite{Mamin01} These parameters fall on the case of $a/b=
0.5$ and $\Lambda/b = 0.005$.

Then, a magnetic moment step, $m_Q$, by adding a single flux quantum, is
numerically estimated as 1.116 $m_b$, where the magnetic moment unit
$m_b \equiv \Phi_0 b/\mu_0$ is $1.65\times 10^{-14} \mathrm{A\cdot
  m^2}$.  The force on the superconducting annulus is given by
\begin{equation}
\label{eq:12}
F = m_Q\frac{dB_{ext}}{dz} \,
\end{equation}
and assuming a modest field gradient, $dB_{ext}/dz$, of 10
T/m,\cite{Gradient} a corresponding force step is estimated to be $1.84
\times 10^{-13}$ N or 184 fN, which causes $\sim 2$ nm static
displacement of a cantilever with $k = 10^{-4}$ N/m.  In principle,
nano- and pico-Newton level can be accessed by stacking these steps.

Practically, a maximum possible force, or the maximum number of steps,
is limited by the critical current of the superconducting material, at
which its superconducting state cannot be sustained. For the Nb annulus
with above geometry, the force limit is roughly estimated to be $\sim$
40 pN from the approximate equation $F_{max} = I_c\pi[(a+b)/2]^2(dB/dz)$
and the critical current density of Nb, $j_c = 9 \times 10^{10} $
A/m$^2$ at $T =$ 6 K.\cite{Geers01}

\paragraph{Force Realization Procedure}
Realization of only a quantum-induced force requires a specially
designed procedure, because the magnetic moment of a superconducting
annulus includes a field-induced component linearly proportional
$B_{ext}$ as seen in Fig. 2.  The magnetic field at the annulus from the
z-gradient magnet, the background, and so on is not simple to null
satisfactorily, considering the small value of $B_0$ ($\sim$ 0.13 Oe for
the parameters above), and would generate a non-zero force offset even
in zero flux quantum state.  Therefore, we need to extract selectively
the flux quantum contribution.

In Fig. 2, the difference of magnetic moment, for example,
between $n = 0$ and $n = 3$ states is always constant,
three-times as large as
$m_Q$ at any fixed $B_{ext}$, and a corresponding change
can be realized by transition between them in principle.
Uncontrolled transitons, such as quantum or thermal tunneling (to be
discussed later), are practically negligible at operation temperatures,
$\sim 4$ K, due to high energy barriers between metastable states.

We suggest a force-realization and cantilever-calibration procedure as
follows: (1) After measuring a zero-reference position of a cantilever
in $n = n_i$ state at $B_{ext}=0$ and $T \ll T_c$,\footnote{The
  $z$-gradient magnet is assumed to be adjusted so that the field at the
  position of the annulus vanishes.} turn on a uniform external field up
to a target value roughly where $n=n_f=n_i+\Delta n$ is a global minimum
of Gibbs free energy.  But, the system still stays at $n=n_i$ state due
to energy barriers.  (2) Then, increase temperature above $T_c$ and
decrease it back to $T \ll T_c$, which would send the system to the
global minimum state, $n=n_f$.  (3) Decrease and turn off the uniform
external field and measure the displacement of the cantilever.  The net
change of magnetic moment, $\Delta n \times m_Q$, is only related to
flux quanta and so is that of a force.  From the number change of flux
quantum ($\Delta n$), annulus dimension, penetration depth, and field
gradient, the change in force can be evaluated and used in determining
the spring constant of the cantilever in combination with its
displacement.

The starting condition $B_{ext}=0$ can be lifted; our procedure is still
valid even for $B_{ext} \neq 0$ as long as it is the same at the
beginning and the end.
This is a great benefit because background magnetic field is
hard to shield and usually disturbs delicate magnetic experiments.

\paragraph{In-Situ Calibration of Field Gradient}
For precise force realization, one has to be able to calibrate the magnetic field gradient $dB_\ext/dz$
as accurate as possible.
Experimentally, it is tricky to know the exact value at the local position
of the annulus on a micro-cantilever, unless the field gradient is strictly uniform  in a large volume
and well pre-calibrated in micro-scale.

\begin{figure}
\includegraphics*[width=7cm]{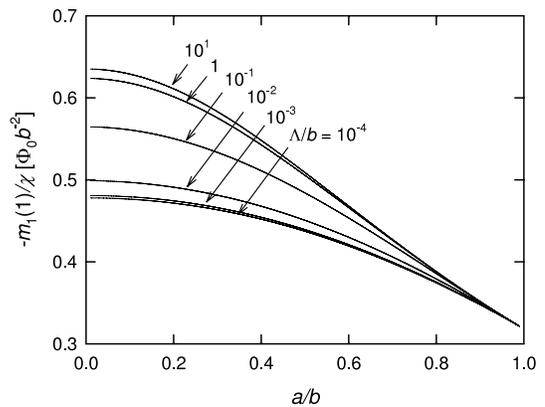}
\caption{$m_1(1)/\chi$ as a function of $a/b$ for different values of
  $\Lambda/b$.}
\label{fig:gradient}
\end{figure}

Here we argue that we can calibrate it in-situ
by means of the characteristic properties of the magnetic moment of the
superconducting annulus.
The magnetization $m(B_\ext)$ in Eq.~(\ref{eq:12}) can be written into the form
\begin{equation}
\label{eq:13}
m = m_1(n) + \chi B_\ext  \,,\quad
\chi \equiv \frac{m_2(B_\ext)}{B_\ext} \,,
\end{equation}
where it should be noticed that $m_1(n)$ and $m_2(B_\ext)$ are linearly
proportional to $n$ and $B_\ext$, respectively.  The field is expanded
as
\begin{equation}
\label{eq:14}
B_\ext(z) = B_{\ext,0} + \frac{dB_\ext}{dz}z \,.
\end{equation}
From Eqs.~(\ref{eq:12}), (\ref{eq:13}), and (\ref{eq:14}),
the force $F(z)$ at the position $z$ is given by
\begin{equation}
\label{eq:15}
F(z) = \left[m_1(1)n + \chi B_{\ext,0}\right]\frac{dB_\ext}{dz}
+ \chi\left(\frac{dB_\ext}{dz}\right)^2z \,.
\end{equation}
Then it follows that
\begin{equation}
\label{eq:16}
\frac{\Delta\omega}{\omega}
= \frac{1}{2}\frac{\Delta k}{k}
= \frac{\chi}{2k}\left(\frac{dB_\ext}{dz}\right)^2 \,,
\end{equation}
where $w$ ($k$) is the natural vibration frequency (Hook's
coefficient) of the cantilever and $\Delta{w}$ ($\Delta{k}$) is the shift
in $w$ ($k$) due to the field gradient. 
Let $\Delta{z}=z_\mathrm{eq}(n)-z_\mathrm{eq}(n=0)$, i.e., the shift in
the equilibrium position $z_\mathrm{eq}$ for non-zero flux $n$.
Since the force $F(z_\mathrm{eq})$ at $z=z_\mathrm{eq}$
in Eq.~(\ref{eq:15}) should be at equilibrium with
$(k+\Delta k)z_\mathrm{eq}$, i.e.,
\begin{equation}
\label{eq:17}
m_1(1)n\frac{dB_\ext}{dz}
= k\Delta z \,,
\end{equation}
one arrives at the relation
\begin{equation}
\label{eq:18}
\frac{dB_\ext}{dz} =
\frac{2n}{\Delta z}\frac{m_1(1)}{\chi}
\frac{\Delta\omega}{\omega} \,.
\end{equation}
We now note that in Eq.~(\ref{eq:18}), the field gradient $dB_\ext/dz$
has been expressed only in known quantities of quantities that can be
determined to high precision: The relative frequency shift
$\Delta\omega/\omega$ is typically probed with high precision.  As shown
in Fig.~\ref{fig:gradient}, the ratio $m_1(1)/\chi$ in Eq.~(\ref{eq:18})
does not depend strongly on the geometry and can be easily determined
accurately.  We therefore conclude that Eq.~(\ref{eq:18}) allows us to
calibrate the field gradient in-situ very precisely.

For $k = 10^{-4}$ N/m and other parameters adopted above for a force
step estimation, a relative frequency shift is estimated to be $2 \times
10^{-3}$ from the Eq.~(\ref{eq:16}), and becomes smaller for lower
spring constant.  The relative shift is much larger than the sharpness
of the resonance peak or 1/Q, with a typical quality factor $Q = 10^4
\sim 10^5$ for single-crystalline Si cantilevers,\cite{Mamin01} and can
be measured to high precision.

\paragraph{Discussion}
The static displacement $\Delta{z}$ of the cantilever can be precisely
measured using, for example, an optic interferometer, which is known to
have the rms noise less than 0.01 nm in 1 kHz bandwidth at low
frequencies.\cite{Rugar89} However, thermal vibration amplitude, $\Delta
z_{th}$, may not be small in comparison with a static displacement,
$\Delta z_{Q}$, due to a single flux quantum.  The ratio, $\Delta
z_{th}/\Delta z_{Q}$, is given by $\sqrt{k k_B T}/F_Q$, where $k_B$ and
$F_Q$ are the Boltzmann constant and the force step due to a flux
quantum, respectively, and is about 0.41 for $k = 10^{-4}$ N/m and $T =
4.2$ K, becoming smaller for decreasing $k$ and $T$.  But, it may not
cause a serious problem because the thermal vibration happens most
dominantly at a resonance frequency ($\sim$ a few kHz), especially for
high-$Q$ cantilever, not disturbing the measurement of static component.
In addition, for large multiples of $F_Q$ the relative error due to
interferometer and thermal noises gets proportionally smaller.


Instead of a superconducting annulus, a superconducting quantum
interference device (SQUID) with leads could be mounted on the
micro-cantilever. Then, instead of thermal cycling, a bias current
through SQUID can be used to permit additional flux quanta, with a
merit that the whole procedure is done at a fixed temperature.
Moreover, each entrance of quantum can be monitored in real time from
the current-voltage characteristics.  For its benefit to be used, an
accurate calculation of magnetic moment for the SQUID case is awaited.

Finally, we briefly remark on the effects of thermal fluctuations.
In principle, the fluxoids trapped in the annulus
may undergo relaxation processes and their number changes.  At our
working temperature ($\sim 4$ K), the relaxation is dominated by
the thermal activation, and governed by the law $\gamma_0\exp(-\Delta
U/k_BT)$, where $\Delta U$ is the energy barrier against the relevant
process.
For very narrow annulus with $b-a\ll\Lambda$, the
underlying mechanism is the phase slip
\cite{Langer67b,McCumber70a,Horane96a,Vodolazov02a}, and the rate is
order of several transitions per second~\cite{Lukens68a,Pedersen01a}.
For a wide annulus with $b-a\gg\Lambda$, the relaxation comes from
crossing of the vortices across the annulus of superconducting thin
film.  Unlike the narrow limit, there is no detailed study of flux-state
transitions in this case, neither experimental nor theoretical.
Naively, the energy barrier can be estimated by the condensation energy
cost inside the vortex core $\Delta{U}\sim H_c^2\xi^2 d/8\pi$, where
$H_c$ is the thermodynamic critical field.  For Nb, $H_c\sim
0.1$ T, and $\Delta{U}\sim 10^5$ K.\cite{Pool00}  It implies that the
effects of the flux relaxation can be safely ignored in our scheme using
wide enough annulus.

\paragraph{Conclusion}
In summary, we suggested pico-Newton force realization based on magnetic
flux quantum in a superconducting annulus.  A stepwise force can be
generated from the annulus on a micro-cantilever in magnetic field
gradient and the step size depends only on fundamental constants and the
length dimensions of the annulus beside the field gradient.  For a 50
nm-thick Nb annulus with inner and outer radii of 5 $\mu$m and 10
$\mu$m, respectively, a force can be created up to $\sim$ 40 pN by a
step of 184 fN, assuming the field gradient of 10 T/m.

\begin{acknowledgments}
M.-S.C. was was supported by the SRC/ERC (R11-2000-071),
the KRF Grant (KRF-2005-070-C00055), and the SK Fund.

\end{acknowledgments}


\begin{thebibliography}{10}
\bibitem{Gradient}
This value is comparable to the typical gradient by  commercial
superconducting or permanent gradient magnets ($2 \sim 20$ T/m).
\end{thebibliography}


\end{document}